\documentstyle[12pt,epsf,epsfig]{article}
\def\be{\begin{equation}}
\def\ee{\end{equation}}
\def\ba{\begin{eqnarray}}
\def\ea{\end{eqnarray}}

\def\bq{\begin{quote}}
\def\eq{\end{quote}}

 at 10truept

\newcommand{\beq}{\begin{equation}}
\newcommand{\eeq}{\end{equation}}
\newcommand{\beqa}{\begin{eqnarray}}
\newcommand{\eeqa}{\end{eqnarray}}

\def\ltap{\ \raise.3ex\hbox{$<$\kern-.75em\lower1ex\hbox{$\sim$}}\ }
\def\gtap{\ \raise.3ex\hbox{$>$\kern-.75em\lower1ex\hbox{$\sim$}}\ }
\def\gl{\ \raise.5ex\hbox{$>$}\kern-.8em\lower.5ex\hbox{$<$}\ }
\def\roughly#1{\raise.3ex\hbox{$#1$\kern-.75em\lower1ex\hbox{$\sim$}}}

\begin{document}

\thispagestyle{empty}
\begin{flushright}
arXiv:0803.2242 [hep-th]\\ 
March 2008
\end{flushright}
\vspace*{1cm}
\begin{center}
{\Large \bf New exact solutions on the Randall-Sundrum 2-brane:}\\ 
\vspace{0.2cm}
{\Large \bf lumps of dark radiation and accelerated black holes}\\
\vspace*{1.5cm} {\large Mohamed Anber\footnote{\tt
manber@physics.umass.edu} and
Lorenzo Sorbo\footnote{\tt sorbo@physics.umass.edu}}\\
\vspace{.5cm} {\em Department of Physics,
University of Massachusetts, Amherst, MA 01003}\\
\vspace{.15cm} \vspace{1.5cm} ABSTRACT
\end{center}
We provide the most general embedding of a purely tensional 2-brane in a 3+1 dimensional bulk described by the AdS C-metric. The AdS C-metric has been first considered as bulk metric by Emparan, Horowitz and Myers~\cite{Emparan:1999wa,Emparan:1999fd}, who have found  metrics describing a brane localized black hole. In~\cite{Emparan:1999wa,Emparan:1999fd}, one of the parameters of the bulk C-metric was fine-tuned to the brane tension. We relax this fine tuning and we find two new classes of solutions, the first describing a time dependent, rotationally symmetric metric, the second describing accelerated black holes on the brane. This is the first exact solution on the brane describing two objects in interaction. We discuss the qualitative CFT interpretation of these solutions.
\vskip2.5cm

\vfill \setcounter{page}{0} \setcounter{footnote}{0}
\newpage
\section{Introduction}

Since the original formulation of the Randall-Sundrum model~\cite{Randall:1999ee}, intensive activity has aimed at the construction of solutions associated to distributions of matter localized on the brane (see e.g. \cite{Chamblin:1999by} - \cite{Grisa:2007qv}). Despite these efforts, the existence of brane localized black holes still represents in general an unsolved question. This problem is made especially interesting by the conjecture~\cite{Tanaka:2002rb, Emparan:2002px} that solutions on the Randall-Sundrum brane should correspond to quantum corrected metrics in the presence of a strongly interacting Conformal Field Theory (CFT). Such a conjecture has allowed~\cite{Tanaka:2002rb,Emparan:2002px} to argue why it has been impossible to find an asymptotically flat, regular and {\em static} black hole localized on the 3-brane. In the dual picture, indeed, such a black hole should receive the quantum corrections associated to the CFT degrees of freedom. Such corrections would induce black hole evaporation, so that the solution cannot be static.

While in the case of a 3-brane embedded in a five-dimensional anti-de Sitter ($AdS_5$) bulk no well behaved black hole solution has been found, in the lower dimensional case of a 2-brane embedded in $AdS_4$ bulk a class of brane-localized black hole metrics was discovered already in 1999 by Emparan, Horowitz and Myers (EHM)~\cite{Emparan:1999wa,Emparan:1999fd}. 

The metrics found in~\cite{Emparan:1999wa,Emparan:1999fd} have allowed to check the conjecture according to which brane black holes should correspond to quantum corrected solutions. Indeed, in~\cite{Tanaka:2002rb,Emparan:2002px} these solutions have been interpreted as quantum corrected conical singularities. More specifically, the fact that the EHM solutions are dressed by a horizon while in 2+1-dimensional gravity we should get naked conical singularities has been interpreted as the effect of a "quantum cosmic censorship"~\cite{Emparan:2002px} for conical singularities: in the presence of a conical singularity, the Casimir effect excites the CFT, that in turn dresses the singularity with horizon.

The construction of EHM can be understood as follows. The Randall-Sundrum brane does not follow a geodesic of the AdS bulk. On the contrary, it experiences a constant acceleration~\cite{Kaloper:1999sm}. The acceleration is determined by the brane tension, i.e. by the effective cosmological constant felt by a brane observer. Therefore a way of finding black hole solutions on the brane is to look for a metric describing an accelerated black hole in the AdS bulk, to cut it with a brane and to impose that the acceleration of the black hole coincides with that of the brane. Now, a metric (called {\em AdS C-metric}~\cite{Plebanski:1976gy})  describing accelerated black holes in {\em four dimensional}\footnote{See~\cite{Charmousis:2003wm} for a study of similar constructions in more than four dimensions.} Anti-de Sitter space is known. By appropriately cutting the AdS C-metric with a brane, EHM could find black hole solutions localized on the 2-brane.

The AdS C-metric depends on four parameters, associated (at least in some limit) to the four dimensional Newton constant, the $AdS_4$ radius, the mass of the black hole and its acceleration. Once we take into account the brane tension, the whole system is characterized by five parameters. In order to stick the black hole onto the brane, EHM impose a relation between the acceleration of the black hole and the tension of the brane, so that the induced metric depends on the four parameters, that can be associated to the effective three dimensional Planck mass, the number of CFT degrees of freedom, the value of the cosmological constant and the black hole mass.

In this paper we will explore the possibility of detuning the brane tension from the bulk black  hole acceleration. As we show in the appendix, all the possible brane configurations  - for a given bulk AdS C-metric -- reduce to two classes. 

The first class has time-dependent induced metrics, and its CFT dual generally describes an evolving lump of radiation, possibly on the top of a conical geometry. Depending on the parameters of the theory, the radiation energy density can stay always finite or can become infinite. In special cases it is possible to find solutions (already discussed in~\cite{Vaganov:2007at}) where the lump of radiation is static.

The second class of solutions is continuously connected to the black hole metric of EHM and leads to static metrics. Such metrics describe in general a pair of  EHM black holes  accelerated by a strut stretched between them (or by two strings pulling them towards infinity).   In the CFT picture, the energy per unit length of this strut has both  classical and  quantum contributions. Indeed, while in pure 2+1 dimensional gravity point particles do not interact, quantum effects~\cite{Soleng:1993yh} generate a force between the particles. The tension of the strut takes into account both contributions. The geometrical interpretation of this solution is quite straightforward, since it turns out that the metric induced on the brane is just a section of the four dimensional C-metric~\cite{Plebanski:1976gy, Kinnersley:1970zw}. This section includes the singularities of the four-dimensional C-metric and therefore describes two accelerated black holes (the two black holes reduce  to a single one  in the case of AdS background and small acceleration).

Our paper is organized as follows. In section 2 we introduce the bulk metric we work with. In section 3 we present the first class of brane embeddings, where the location of the brane in the bulk (as well as the induced metric) is time dependent. In section 4 we present the class of time independent embeddings. In section 5 we discuss our results - especially in relation to the conjecture of~\cite{Tanaka:2002rb, Emparan:2002px} - and we draw our conclusions. The appendix contains the proof that the solutions described in this paper represent the unique possible embeddings of a purely tensional brane in a AdS C-metric bulk. 

\section{The setting}

The AdS C-metric is given by
\begin{equation}\label{cmetric}
ds^2=\frac{1}{A^2(x-y)^2}\left[-H(y)\,dt^2+\frac{dy^2}{H(y)}+\frac{dx^2}{G(x)}+G(x)\,d\phi^2 \right]\,\,,
\end{equation}
where the functions $H\left(y\right)$ and $G\left(x\right)$ are
\begin{eqnarray}
H(y)&=&\lambda-k\,y^2+2\,G_4\,mA\,y^3\,\,,\nonumber\\
G(x)&=&1+k\,x^2-2\,G_4\,mA\,x^3\,\,.
\end{eqnarray}

In  the expressions above, $\lambda >-1$ and $k=-1\,,0\,,+1$, while $G_4$ denotes the four dimensional Newton constant. This metric describes one or two accelerated black holes on an Anti de Sitter background with radius $\ell_4=1/A\sqrt{\lambda+1}$. Zeros of the function $H\left(y\right)$ corresponds either to black hole or acceleration horizons. In particular, for $k=-1$, $G_4 \,mA<1/3\sqrt{3}$ and $\lambda\ge 0$ (i.e. $A\le 1/\ell_4$), the metric~(\ref{cmetric}) describes a black hole of mass $m$, subject to acceleration $A$ and with a horizon with spherical topology. The black hole is accelerated by a string that pulls it towards the AdS boundary. For $k=-1$, $G_4 \,mA<1/3\sqrt{3}$ and $-1<\lambda<0$ the metric describes {\em two} accelerated black holes in AdS space. A strut between the black holes prevents them from coalescing (alternatively, the same effect is achieved by two strings that pull the black holes towards the AdS boundary). This situation is qualitatively similar to that of the usual C-metric on a Minkowskian background, that corresponds to the case $\lambda=-1$.  For $k=0$ the metric describes an accelerated version of the AdS planar black hole, whereas for $k=+1$ it describes an accelerated AdS black hole with hyperbolic horizon\footnote{The majority of the literature on the AdS C-metric focuses on the case $k=-1,\,G_4 \,mA<1/3\sqrt{3}$, where the coordinate $x$ behaves like $\cos\theta$ in polar coordinates. If these conditions are not met, the function $G\left(x\right)$ vanishes only in one point, implying that $x$ is now akin to a radial coordinate and the horizon is noncompact. As a consequence, there is no conical singularity in the metric that can be interpreted as a string pulling the black hole. Indeed, in this case the object that accelerates the black hole is entirely hidden behind  the horizon.}. More detailed descriptions of the AdS C-metric can be found in~\cite{Podolsky:2002nk,Dias:2002mi,Krtous:2005ej}. For the present work, all we need to know is that $-1/y$ is a radial coordinate from the particle (that is located at a singularity at $y\rightarrow -\infty$). The coordinate $x$ can be roughly interpreted (for $k=-1$ and $G_4\,mA\le 1/3\sqrt{3}$) as $\cos\theta$ in polar coordinates. The $x$ coordinate is bound to be larger than $y$ ($x>y$), and the surface $x=y$ corresponds to the $AdS_4$ boundary.  

We now want to embed a brane with tension $\tau\equiv\left(1+\delta\right)/\left(2\pi\,G_4\ell_4\right)$ in this bulk. The quantity $\delta$ is defined in such a way that $\delta=0$ corresponds to a critical tension brane, that in absence of bulk matter ($m=0$) would lead to a Minkowskian brane induced metric.

As we show in the appendix, the most general embedding of a purely tensional $Z_2$-symmetric 2-brane in such a bulk is given either\footnote{The fact that possible embeddings come in pairs should not come as a surprise, given the invariance of the metric~(\ref{cmetric}) under the exchange $t\leftrightarrow i\,\phi$, $x\leftrightarrow y$, $H\leftrightarrow G$.} by $y=\psi\left(t\right)$ or $x=\xi\left(\phi\right)$. In the next section we will consider the first situation, while the case $x=\xi\left(\phi\right)$ will be discussed in section 4. 

\section{Time dependent solutions}

The first class of embeddings we consider are of the form $y=\psi\left(t\right)$. The function $\psi\left(t\right)$ is determined by Israel's junction condition, and obeys the differential equation
\begin{equation}\label{eqpsit}
\left(\frac{d\psi}{dt}\right)^2=H(\psi\left(t\right))^2-\frac{H(\psi\left(t\right))^3}{\alpha^2}\,,
\end{equation}
where we have defined the quantity
\begin{equation}
\alpha\equiv\left(1+\delta\right)\,\sqrt{1+\lambda}=2\pi\,G_4\,\tau/A\,\,,
\end{equation}
and $\alpha=1$  corresponds to the case studied by EHM. In addition to (\ref{eqpsit}), the junction conditions give the auxiliary equation
\begin{equation}\label{eqpsitt}
2H(\psi)\frac{d^2\psi}{dt^2}-2H'(\psi)(\frac{d\psi}{dt})^2+H'(\psi)H^2(\psi)=0\,.
\end{equation}

Using equation~(\ref{eqpsit}), and using $y$ rather than $t$ as independent variable\footnote{This is possible as long as $d\psi/dt\neq 0$. The special case $\psi=$constant was studied in~\cite{Vaganov:2007at}.}, we find the induced metric on the brane
\begin{equation}\label{indy}
ds^2=\frac{1}{A^2\,\left(x-y\right)^2}\,\left[-\frac{dy^2}{\alpha^2-H\left(y\right)}+\frac{dx^2}{G\left(x\right)}+G\left(x\right)\,d\phi^2\right]\,.
\end{equation}

The dynamics of this system can be made more transparent by performing the change of variable $y=-1/Ar$ and subsequently defining $r=a\left(\eta\right)$ where $a\left(\eta\right)$ obeys the Friedmann-like equation
\begin{equation}\label{cosmoeq}
\frac{a'\left(\eta\right)^2}{a\left(\eta\right)^4}=A^2\left(\alpha^2-\lambda\right)+\frac{k}{a\left(\eta\right)^2}+\frac{2\,G_4\,m}{a\left(\eta\right)^3}
\end{equation}
describing a 2+1 dimensional cosmology with closed, flat, or open slices (depending on the value of $k$) whose matter content is given by a cosmological constant $\propto  A^2\left(\alpha^2-\lambda\right)/G_3$ and radiation with temperature $\propto \left(G_4\,m/G_3\right)^{1/3}/a\left(\eta\right)$. In terms of the variables $\eta$, $x$ and $\phi$ the metric reads
\begin{equation}\label{cosmomet}
ds^2=\frac{a\left(\eta\right)^2}{\left(1+A\,x\,a\left(\eta\right)\right)^2}\,\left[-d\eta^2+\frac{dx^2}{\left(1+k\,x^2-2\,G_4m\,A\,x^3\right)}+\left(1+k\,x^2-2\,G_4m\,A\,x^3\right)\,d\phi^2\right]\,.
\end{equation}

In these coordinates the limit $A\rightarrow 0$ is straightforward and the resulting geometry describes a cosmological metric filled with (dark) radiation~\cite{Kraus:1999it} - \cite{Bowcock:2000cq}.

In general, the metric~(\ref{cosmomet}) describes a lump of radiation on a background with cosmological constant $A^2\,\left(\alpha^2-\lambda-1\right)=\left(2\,\delta+\delta^2\right)/\ell_4^2$, as one can see by writing the stress energy tensor that supports the metric~(\ref{cosmomet})
\begin{equation}\label{tmunuy}
T^\mu{}_\nu=-\frac{A^2}{8\pi G_3}\,\left(\alpha^2-\lambda-1\right)\,{\mathrm {diag}}\left(1,\,1,\,1\right)+\frac{G_4m}{8\pi G_3\,a\left(\eta\right)^3}\,\left(1+A\,x\,a\left(\eta\right)\right)^3\,{\mathrm {diag}}\left(-2,\,1,\,1\right)\,,
\end{equation}
where $G_3$ is the three dimensional Newton constant. As we will see in detail in the next subsections, the size of the lump of radiation, as well as the time-scale over which it evolves, are of the order of $A^{-1}$. Note that the cosmological constant that appears in eq.~(\ref{tmunuy}) does not have the same value as the first term on the right hand side of the "Friedmann equation"~(\ref{cosmoeq}). 

In the following subsections we will discuss this metric for some representative choices of parameters. We will focus on the case of a brane with critical tension ($\delta=0$, $\alpha^2=1+\lambda$), where the interpretation of the metric is the most transparent.

Before discussing these special cases, let us remark that by appropriately tuning the parameters of this class of solutions it is possible to obtain static configurations. These configurations can be obtained both for a subcritical and for a critical brane, and describe static, self-gravitating lumps of radiation. A detailed description of these specific cases can be found in~\cite{Vaganov:2007at}.

\subsection{Critical, closed slices ($\delta=0$, $k=-1$)}

For $k=-1$ and $mA<1/3\sqrt{3}$ the function $G\left(x\right)$ has three zeros, that we denote as $x_0,\,x_1,\,x_2$ with $x_0<x_1<0<x_2$. $G$ is positive for $x<x_0$ and for $x_1<x<x_2$. In the latter range, we interpret $x$ roughly as $\cos\theta$ in polar coordinates. $x=x_1$ and $x=x_2$ then correspond to the polar axis. Since $\left|G'\left(x_1\right)\right|\neq\left|G'\left(x_2\right)\right|$ there is a conical singularity either at the north or at the south pole. We redefine $\phi$ so that the axis $x=x_2$ is regular, that corresponds to having a deficit angle along $x=x_1$. Such a deficit angle is interpreted as due to a string responsible for the acceleration of the black hole. 

For $\delta=0$, $k=-1$, the brane induced metric~(\ref{indy}) takes the form
\begin{eqnarray}\label{indy32}
ds^2=&&\frac{1}{A^2\left(x-y\right)^2}\,\left[-\frac{dy^2}{1-y^2-2\,G_4mA\,y^3}+\frac{dx^2}{1-x^2-2\,G_4mA\,x^3}+\right.\nonumber\\
&&+\left.\left(1-x^2-2\,G_4mA\,x^3\right)\,d\phi^2\right]\,\,.
\end{eqnarray}

In this section we will study the limit $G_4mA\ll 1$. To start with, we note that, for $m=0$, eq.~(\ref{indy32}) reduces to
\begin{equation}
ds^2=\frac{1}{A^2\left(x-y\right)^2}\,\left[-\frac{dy^2}{1-y^2}+\frac{dx^2}{1-x^2}+\left(1-x^2\right)d\phi^2\right]\,\,,
\end{equation}
that is Minkowski space in disguise, since the transformation
\begin{eqnarray}
&&x\left(t,\,r\right)=\frac{A^2t^2-A^2r^2+1}{\sqrt{4\,A^2r^2+A^4\left(t^2-r^2+1/A^2\right)^2}}\,\,,\nonumber\\
&&y\left(t,\,r\right)=\frac{A^2t^2-A^2r^2-1}{\sqrt{4\,A^2r^2+A^4\left(t^2-r^2+1/A^2\right)^2}}\,\,,
\end{eqnarray}
brings it to the form $ds^2=-dt^2+dr^2+r^2 d\phi^2$. This transformation helps to clarify the evolution of the lump of radiation associated to the stress energy tensor~(\ref{tmunuy}). At first order in $G_4mA$, we get indeed that the energy distribution supporting our solution is given, in terms of the coordinates $t$ and $r$, by
\begin{equation}
\rho\left(t,\,r\right)=T_{tt}=\frac{2\,G_4\,m}{\pi G_3\,A^3}\frac{r^4+\left(t^2+1/A^{2}\right)^2+r^2\,\left(4\,t^2+2/A^{2}\right)}{\left[4\,r^2/A^{2}+\left(t^2-r^2+1/A^{2}\right)^2\right]^{5/2}}+{\cal {O}}\left(m^2\right)\,\,.
\end{equation}

\begin{figure}[htp]
\centering
\includegraphics[width=0.5\textwidth]{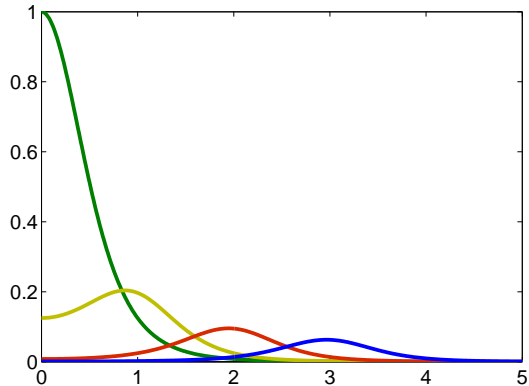}
\caption{The function $\rho\left(t,\,r\right)$ for different values of $t=0,\,1,\,2,\,3$. Here $2G_4m/\pi G_3A^3=A=1$. Each curve has a maximum at $r=t$.}
\end{figure}

We plot the profile of $\rho\left(r\right)$ at different times in figure 1. This shows  that the solution describes a circular shell of radiation that contracts for $t<0$, reaches a maximal energy density at $r=0$ when $t=0$ and then bounces to infinity. When $t=0$, $\rho\left(t=0,\,r\right)\propto \left(r^2+1/A^2\right)^{-3}$. To first order in $G_4mA$, it is possible to compute the total mass of the lump as $2\pi\int \rho\left(t,\,r\right)\,r\,dr=\left(G_4mA/G_3\right)\,\left(1+{\cal {O}}\left(G_4mA\right)\right)$. Note that this shell moves on the top of a conical geometry with deficit angle $4\pi G_4 m A\,\left(1+{\cal {O}}\left(G_4mA\right)\right)$, corresponding to a mass $\left(G_4 m A/2G_3\right)\left(1+{\cal {O}}\left(G_4mA\right)\right)$ located at the origin of the system \cite{Emparan:1999wa}.

At variance with the solution of EHM, the metric~(\ref{indy32}) does not display a horizon on the brane. From the CFT point of view, the quantum cosmic censorship of conical singularities seems not to be at work for this state. On the other hand it is also worth noting that this is a time-dependent solution, whereas the censored solution of EHM was static.

\subsection{Critical, flat slices ($\delta=0$, $k=0$)}

In this case the induced metric reads 
\begin{equation}\label{k0delta0y}
ds^2=\frac{1}{A^2\left(x-y\right)^2}\,\left[-\frac{dy^2}{1-2\,G_4mA\,y^3}+\frac{dx^2}{1-2\,G_4mA\,x^3}+\left(1-2\,G_4mA\,x^3\right)d\phi^2\right]\,\,,
\end{equation}
and the variable $x$ ranges between $-\infty<x<\left(2\,G_4mA\right)^{-1/3}$. $x=x_m\equiv\left(2\,G_4mA\right)^{-1/3}$ corresponds to the origin of polar coordinates. We avoid a conical singularity by giving the angle $\phi$ a periodicity $0<\phi<4 \pi x_m/3$. In the limit $G_4mA\rightarrow 0$, this period diverges and $\phi$ becomes a linear coordinate.

This brane induced metric~(\ref{k0delta0y}), like the one described in the previous subsection, describes an evolving lump of radiation. However, its properties are different. Let us start by looking at the center of the distribution of radiation, $x\simeq x_m$. In this region it is convenient to use the "cosmological" metric~(\ref{cosmomet}). We have $\rho\left(\bar{t}\right)=G_4\,m\,M_3\,\left(1+A\,a\left(\bar{t}\right)\,x_m\right)^3/8\pi G_3 a\left(\bar{t}\right)^3$, where we have switched from the "conformal time" $\eta$ to "physical time" $\bar{t}$, i.e. $d\eta=d\bar{t}/a\left(\bar{t}\right)$. The function $a\left(\bar{t}\right)$ is obtained by solving the Friedmann-like equation $\dot{a}^2/a^2=A^2+2G_4m/a^3$. In the early time regime $a^3\ll 2G_4m/A^2$, the cosmology is radiation dominated, $a\left(\bar{t}\right)\propto m^{1/3}\,\bar{t}^{2/3}$. In this case the center of our distribution experiences a "big bang" with infinite energy density as $\bar{t}\rightarrow 0$ (i.e.  $y\rightarrow -\infty$). This is different from the situation considered in the previous subsection where the energy density was always finite.

In the opposite limit $a\left(\bar{t}\right)\gg \left(2\,G_4mA\right)^{1/3}/A=\left(A\,x_m\right)^{-1}$, the metric~(\ref{k0delta0y}) reduces, close to the origin $x\simeq x_m$, to Minkowski metric modulo an overall scaling of the coordinates (this can be seen most clearly by considering the metric in its form~(\ref{cosmomet})). In the same limit, the stress energy tensor for brane radiation goes to the constant value $T^\mu{}_\nu\simeq A^2/(16\pi G_3)\,{\mathrm {diag}}\left(-2,1,1\right)$. 

In order to understand the behavior of the system far from $x=x_m$ let us consider the limit $m\rightarrow 0$, with $x$ finite. In this limit the brane is actually flat, and $\phi$ becomes a linear coordinate. This is shown explicitly by the fact that for $m=0$ the metric~(\ref{k0delta0y}) reduces to 
\begin{equation}
ds^2=\frac{1}{A^2\left(x-y\right)^2}\,\left(-dy^2+dx^2+d\phi^2\right)
\end{equation}
that is brought to the Minkowskian form $ds^2=-dT^2+dX^2+dY^2$ by the transformation
\begin{eqnarray}
x-y&=&\frac{1}{A\left(T+X\right)}\,\,,\nonumber\\
x+y&=&A\left[\left(T-X\right)-\frac{Y^2}{T+X}\right]\,\,,\nonumber\\
\phi&=&\frac{Y}{T+X}\,\,.
\end{eqnarray}

Since, for $m=0$, $\phi$ is a linear coordinate, the limit of small $m$ and finite $x$ will correspond to the regime where $\phi$ is "almost" linear, i.e. far from the center of our distribution. As a consequence, we can compute the energy density $\rho\left(y,\,x,\,\phi\right)$ far from the center of the distribution of radiation by considering its expression at first order in $m$. Making use of the rotational symmetry of the system, we can set $\phi=Y=0$. Then, the energy density of our fluid is given for $G_4mA\ll 1$ by 
\begin{equation}
\rho\left(T,\,X,\,Y\right)=\frac{G_4\,m\,A^3}{8\pi\,G_3}\,\frac{3 A^4\,(X+T)^4+2\,A^2\,(X+T)^2+3 }{4\,A^5\,(T+X)^5}+{\cal {O}}\left(\left(G_4mA\right)^2\right)\,\,,
\end{equation}
that, for large $T$, decreases as $T^{-1}$. 

To sum up, in the case $k=0$ our system describes a circular lump of radiation that starts from infinite density at its center and relaxes down  to Minkowski space at large times. 

Note that in  this case the origin is regular provided we choose the right periodicity for $\phi$, that implies that -- differently from the case considered in the previous section -- the matter content here is just that of the lump of (dark) radiation, and there is no localized matter on the brane.

\section{Time independent solutions}

In the second class of solutions the brane embedding is given by  $x=\xi(\phi)$. Using the $K_{11}$ component in Isreal junction conditions~(\ref{extrinsic curvature}) we find that $\xi(\phi)$ satisfies the differential equation 
\begin{equation}\label{main eq BH}
\left(\frac{d\xi}{d\phi}\right)^2=\frac{G(\xi(\phi))^3}{\alpha^2}-G(\xi(\phi))^2\,.
\end{equation}
Also, from the $K_{33}$ component of~(\ref{extrinsic curvature}) we find the auxiliary equation
\begin{equation}\label{consistency eq}
2\,G^2(\xi)\,\frac{d^2\xi}{d\phi^2}-3\,G(\xi)\,G'(\xi)\,\left(\frac{d\xi}{d\phi}\right)^2-G^3(\xi)\,G'(\xi)=0\,.
\end{equation}

In this section we will consider only the case $k=-1$, that corresponds to the situation  most thoroughly studied in the literature. Since for $k=-1$ we have that $G\left(x\right)\le 1$ in the interval $x_1<x<x_2$, eq.~(\ref{main eq BH}) can be solved only if $\alpha^2<1$. We will therefore assume $\alpha^2<1$ from now on.

Using $x$ as independent variable\footnote{The case $x=$constant (with $x$ obtained by looking for zeros of $G'\left(x\right)$) has been studied by EHM.} we find the induced metric
\begin{equation}\label{induced metric on xi}
ds^2=\frac{1}{A^2(x-y)^2}\left[-H(y)\,dt^2+\frac{dy^2}{H(y)}+\frac{dx^2}{G(x)-\alpha^2} \right]\,.
\end{equation}
To understand this metric we perform the following transformations
\begin{eqnarray}
\nonumber
x^{\prime}&=&\frac{x}{\sqrt{1-\alpha^2}}\,,\qquad y^{\prime}=\frac{y}{\sqrt{1-\alpha^2}}\,,\qquad t^{\prime}=\sqrt{1-\alpha^2}\,t\\
A^{\prime}&=&A\sqrt{1-\alpha^2}\,,\qquad\lambda^{\prime}=\frac{\lambda}{1-\alpha^2}\,,\label{transformation of induced BH}
\end{eqnarray} 
that yield
\begin{equation}\label{induced metric for BH}
ds^2=\frac{1}{A^{\prime\,2}(x^{\prime}-y^{\prime})^2}\left[- H(y^{\prime})\,dt^{\prime\,2}+\frac{dy^{\prime\,2}}{H(y^{\prime})}+\frac{dx^{\prime2}}{G(x^{\prime})}\right]\,\,,
\end{equation} 
where $H(y^{\prime})=\lambda^{\prime}-k\,y^{\prime\,2}+2\,G_{4}\,m\,A^{\prime}\,y^{\prime\,3}$ and $G(x^{\prime})=1+k\,x^{\prime\,2}-2\,G_{4}\,m\,A^{\prime}\,x^{\prime\,3}$. The Ricci scalar of the metric~(\ref{induced metric for BH}) is a constant $R=-6\,A^{\prime\,2}(1+\lambda^{\prime})$.

The metric~(\ref{induced metric for BH}) is a constant $\phi$ section of the C-metric~\cite{Plebanski:1976gy} (on a Minkowski, de Sitter, anti-de Sitter background, depending on the value of $\lambda^\prime$). Therefore, it describes accelerated black holes in 2+1 dimensions on the background of CFT matter and of a cosmological constant given by $\Lambda_{eff}=-A^{\prime\,2}(1+\lambda^{\prime})$, as can be seen by writing the stress energy tensor induced on the brane
\begin{equation}\label{energy momentum tensor for the induced BH}
T^{\prime\, \mu}_{\nu}=	\frac{A^{\prime\,2}}{8\pi G_3}\,(1+\lambda^{\prime})\,\mbox{diag}(1,1,1)+\frac{G_{4}\,m\,A^{\prime\,3}}{8\pi G_3}\,(x^{\prime}-y^{\prime})^3\,\mbox{diag}(1,1,-2)\,.
\end{equation}  

\begin{figure}[htp]
\centering
\includegraphics[width=0.4\textwidth]{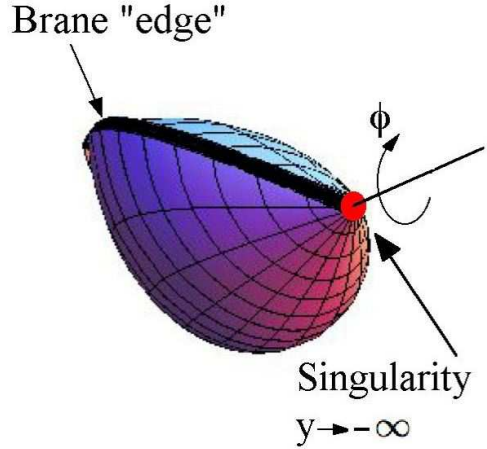}
\caption{Schematic plot of the embedding of the brane defined by $x=\xi\left(\phi\right)$, where the function $\xi\left(\phi\right)$ is a solution of eq.~(\ref{main eq BH}).\label{edge}}
\end{figure}

Let us now study the geometry of this system. The bulk black hole horizon has spherical topology provided $G_4\,m\,A<1/3\sqrt{3}$. Using the transformation in (\ref{transformation of induced BH}) we see immediately that this implies that $G_4\,m\,A^{\prime}<1/3\sqrt{3}$ and, as we show below, the brane $x=\xi(\phi)$ cuts the bulk such that the horizon in (\ref{induced metric for BH}) has circular topology. For $G_4\,m\,A>1/3\sqrt{3}$ the bulk black hole horizon has $R^2$ topology (i.e. the black hole horizon extends all the way to the boundary of $AdS_4$). In this case, depending on $\alpha$, we have either $G_4\,m\,A^{\prime}<1/3\sqrt{3}$ or $G_4\,m\,A^{\prime}>1/3\sqrt{3}$ which corresponds to having a brane black hole horizon with circular ($S^1$) or $R^1$ topology, respectively. This horizon is located at the smallest zero of $H(y^{\prime})$, and dresses a singularity at $y=-\infty$.   

For a critical brane with $\delta=0$ we have $1-\alpha^2=-\lambda$. This  corresponds to  $\lambda^{\prime}=-1$, so that the effective cosmological constant on the brane vanishes. Hence the induced metric in (\ref{induced metric for BH}) degenerates to a constant $\phi$ section of the C-metric  which describes a pair of black holes accelerating in asymptotically flat spacetime \cite{Kinnersley:1970zw}.

For a subcritical brane, $-1<\delta<0$, we have $0<\lambda<\infty$, and therefore $-1<\lambda^{\prime}<\infty$. In this situation we obtain a negative cosmological constant on the brane, and the metric (\ref{induced metric for BH}) describes a constant $\phi$ section in the $AdS_4$-C metric \cite{Plebanski:1976gy}.

Finally, for a supercritical brane with $\delta>0$ we obtain $-\infty<\lambda^{\prime}<-1$ and hence a positive cosmological constant on the brane. This metric is a constant $\phi$ section in the $dS_4$-C metric which describes an accelerated pair of black holes in $dS_4$ space \cite{Mann:1995vb, Podolsky:2000pp}. 

Now we turn to the discussion of the embedding of the brane in the bulk. We first consider the case $m=0$ which corresponds to a empty $AdS_4$ bulk. In this case one can readily integrate eq.~(\ref{main eq BH}) to obtain
\begin{equation}\label{solution at m=0}
\xi_{m=0}(\phi)=\frac{\sqrt{1-\alpha^2}\,\sin\phi}{\sqrt{\alpha^2\,\cos^2\phi+\sin^2\phi}}\,.
\end{equation}
It is clear from the solution that the function $\xi_{m=0}\left(\phi\right)$ is periodic and that its period  matches that of the angle $\phi$ of the bulk, i.e. $\Delta\phi_{brane}=\Delta\phi_{bulk}=2\,\pi$.  
The above solution, obtained for $m=0$, can be used to explain the topology of the constant-$y$ surfaces (and therefore of the horizon of induced black hole) for small values of  $m$. To this end consider a unit $S^2$ sphere given by the embedding $\xi=\cos\theta$, $Y=\sin\theta\cos\phi$ and $Z=\sin\theta\sin\phi$. One can see immediately that the above equation (\ref{solution at m=0}) describes the plane $Z=\alpha\xi/\sqrt{1-\alpha^2}$ intersecting the given sphere in a circle. 

\begin{figure}[htp]
\centering
\includegraphics[width=0.5\textwidth]{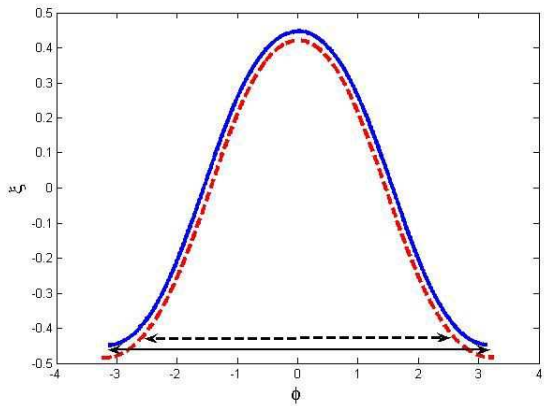}
\caption{Numerical solution of eq.~(\ref {main eq BH}) for the case $k=-1$ using a critical brane $\delta=0$. We take $\lambda=-0.2$, and $G_4mA=0.0$ and $0.15$ for the solid and dashed lines respectively. The arrows on the figure indicate the periodicity of $\phi$. In the first case the period of $\xi\left(\phi\right)$ of the brane embedding coincides with that of the bulk: $\Delta\phi_{\mbox{\scriptsize bulk}}=\Delta\phi_{\mbox{\scriptsize brane}}=2\pi$. In the second case we see that the period of the brane embedding is larger than the periodicity of $\phi$.\label{the period on the brane}}
\end{figure}

Let us then consider the case $G_4m\,A<1/3\sqrt{3}$, with nonvanishing $m$. In this case  $G(x)$ vanishes at $x=x_1\,,x_2$ where $x_1<x_2$, these directions correspond to the axis of rotation. To avoid a conical singularity at $x=x_2$, we take $\phi$ to have the period $\Delta\phi_{bulk}=4\,\pi/|G^{\prime}(x_2)|$. Since we have adjusted the period at $x=x_2$, one can no longer adjust the period at $x=x_1$ and we encounter a conical singularity along this axis. For $m\neq 0$ one can not find the solution of eq.~(\ref{main eq BH}) in a closed form. However, numerical integration shows that the solution is periodic and bounded, and the period of the brane embedding is always larger than that of the bulk, i.e. $\Delta\phi_{brane}>\Delta\phi_{bulk}$ as shown in figure~\ref{the period on the brane}. This discrepancy between the two periods indicates the existence of a codimension-one object, an edge, on the brane~\footnote{An exact solution describing a codimension-one object on the brane was first described in~\cite{Gregory:2001xu,Gregory:2001dn}, while its CFT interpretation was studied in~\cite{Grisa:2007qv}. Note that however in our case the defect is bounded by two black holes, whereas the object considered in~\cite{Gregory:2001xu,Gregory:2001dn,Grisa:2007qv}  has infinite extension.}. In figure~\ref{edge} we provide a schematic plot of the embedding of the brane in the bulk. The energy per unit length of this edge is given by~\cite{Hayward:1993my}
\begin{equation}\label{the expresion of the tension}
\mu=-\frac{1}{4\pi\,G_4}\cos^{-1}(n_0^\mu n_{1}{}_\mu)\,,
\end{equation}
where $n_{0}$ and $n_{1}$ are the unit normals on the two sides of the edge. Using eqs. (\ref{main eq BH}) and (\ref{unit normal for x brane}), and imposing the symmetry requirement  $\xi(-\Delta\phi_{\mbox{\scriptsize bulk}}/2)=\xi(\Delta\phi_{\mbox{\scriptsize bulk}}/2)$ and $\xi^{\prime}(-\Delta\phi_{\mbox{\scriptsize bulk}}/2)=-\xi^{\prime}(\Delta\phi_{\mbox{\scriptsize bulk}}/2)$  we obtain
\begin{equation}\label{tension of the edge}
\mu=-\frac{1}{4\pi\,G_4}\cos^{-1}\left[2\,\alpha^2/G(\Delta\phi_{\mbox{\scriptsize bulk}}/2)-1 \right]\,,
\end{equation} 
where $G(\Delta\phi_{\mbox{\scriptsize bulk}}/2)=G(\xi(\Delta\phi_{\mbox{\scriptsize bulk}}/2))$. From eq. (\ref{tension of the edge}) we see that the maximum value of the tension is $\left| G_4\mu_{\mbox{\scriptsize max}}\right|=1/4$. 

 One can also obtain an expression for the energy per unit length $\tau$ of the edge  from the point of view of an observer on the brane. To this end we write the Isreal junction conditions  for the brane-induced metric
\begin{equation}
{\cal K}_{\mu\nu}-\ell_{\mu\nu}{\cal K}=-8\pi G_3 \ell_{\mu\nu}\tau\,,
\end{equation}
where ${\cal K}$ and $\ell$ are respectively the extrinsic curvature and the induced metric on the edge. Using the metric in  (\ref{induced metric on xi}),  and remembering that the angular coordinate $\phi$ ranges between $-\Delta\phi_{\mbox{\scriptsize bulk}}/2$ and $\Delta\phi_{\mbox{\scriptsize bulk}}/2$, we obtain
\begin{equation}\label{tension tau}
\tau=-\frac{A}{4\pi G_3}\sqrt{-\alpha^2+G(\Delta\phi_{\mbox{\scriptsize bulk}}/2)}\,,
\end{equation}
where in general one does not expect to have $\mu=\tau$.

\begin{figure}[htp]
\centering
\includegraphics[width=0.5\textwidth]{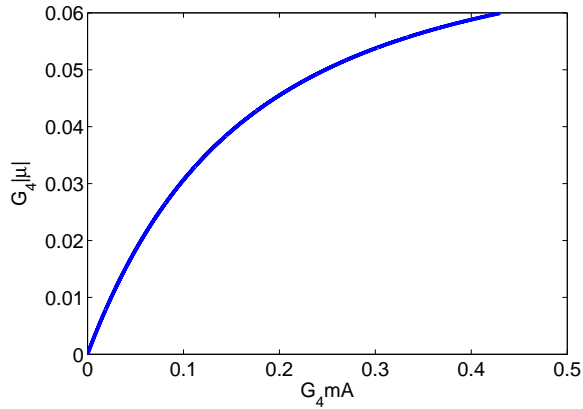}
\caption{The absolute value of the energy per unit length $\left|\mu\right|$ of the edge as a function of $G_4mA$. We take $\lambda=-0.2$ for the case of critical brane $\delta=0$.\label{the tension and the range for k=-1}}
\end{figure}

 In figure~\ref{the tension and the range for k=-1} we plot the absolute value of the energy per unit length for the range $G_4mA'<1/3\sqrt{3}$. For small values of $mA$ one can show by means of numerical techniques that the tension of the edge to a first order in $mA$ is given on the critical brane by the expression
\begin{equation}\label{mu for small mA}
\mu=mA^{\prime}+{\cal O}(G_4m^2A^{\prime\,2})\,,
\end{equation}
where $A^{\prime}=A\sqrt{1-\alpha^2}$ is the acceleration of the brane-induced black hole. In addition, we can express $\tau$ in terms of $\mu$ and use the relation $G_3=G_4/2\ell_4$ to obtain for small values of $mA$ 
\begin{equation}
\tau=mA^{\prime}+{\cal O}(G_4m^2A^{\prime\,2})\,\,. 
\end{equation}
The first term in the above expression is classical in nature and appears due to the fact that we accelerate massive objects. The second term is expected to be different from the ${\cal {O}}\left(G_4m^2A^{\prime\,2}\right)$ correction to $\mu$ in~(\ref{mu for small mA}), and is associated to the CFT correction. Although gravity is dynamically trivial in 2+1 dimensions, the quantum effects generate a force between particles. The existence of ${\cal {O}}\left(G_4\right)$ corrections to the strut tension reflects the presence of such quantum effects.

\section{Discussion and Conclusions}

In this paper we have studied the most general embeddings of a vacuum 2-brane in a AdS C-metric background. Our solutions generalize those found by Emparan, Horowitz and Myers in 1999~\cite{Emparan:1999wa,Emparan:1999fd}, and can be divided into two classes. The first class (studied in section 3) contains time dependent metrics, whose CFT dual describes a rotationally invariant, time dependent lump of radiation. By studying two specific cases we have seen that, depending on the choices of parameters, the radiation can be either in the form of a collapsing and bouncing shell or in the form of a lump that, starting from infinite density at its center, eventually relaxes to a vacuum configuration. The second class of solutions that we have found  describes one (or two) accelerated black holes kept in a static configuration either by a strut or by one (or two) strings.

The class of brane metrics studied in section 4 can be reduced to constant $\phi$ sections of the general C-metric~(\ref{cmetric}). It is straightforward to see in the same way that the solutions of section 3 can be obtained by taking constant $t$ sections of the C-metric~(\ref{cmetric}). In this class of solutions the radial coordinate $y$ turns into a time coordinate in the regions inside the horizon of the full C-metric~(\ref{cmetric}). Already EHM had noticed that their 2+1 dimensional black hole was characterized by the same metric as an equatorial section of a ordinary, 3+1 dimensional Schwarzschild black hole (with a deficit angle). Therefore, all the brane induced metrics explicitly found by cutting a AdS C-metric with a vacuum brane appear to be sections of a four dimensional vacuum metric. One might wonder whether this behavior has any deep origin or it is only accidental.

Let us discuss the CFT interpretation of our solutions. The interpretation of the solutions of section 4 is rather straightforward. In pure 2+1 dimensional gravity two particles do not interact, since lower dimensional gravity is non dynamical. However, the solution of EHM shows that, when dressed with the effects of a CFT, a particle in 2+1 dimensions generates an attractive field. Our solutions of section 4 describe  a pair of such particles accelerated by the presence of a strut (that in 2+1 dimensions is a codimension-1 object). Since these dressed particles attract each other, we need to correct the force of the strut to pull them away from each other: this is precisely what is described by a constant $\phi$ section of a C-metric. It would be interesting to study the CFT counterpart of this solution by computing the quantum corrections to a 2+1 dimensional geometry containing two accelerated conical singularities.

The interpretation of the solutions of section 3 is less straightforward. Clearly, the solution contains an evolving lump of CFT. In the limit of vanishing acceleration $A\rightarrow 0$, the lump of CFT becomes homogeneous and isotropic, and the solution converges to the (dark) radiation dominated cosmology studied for instance in~\cite{Kraus:1999it} - \cite{Bowcock:2000cq}. For nonvanishing values of $A$, on the other hand, homogeneity is lost and only isotropy is maintained. The case $k=-1$ is especially interesting, since in this case the dark radiation evolves on the top of a conical geometry, implying that the dual description of this solution contains both dark radiation and a (pointlike) particle. At variance with the static solution of EHM, this conical singularity is naked, even if it is surrounded by an evolving bath of radiation. Therefore it looks that in this case the "quantum censorship of conical singularities" invoked in~\cite{Emparan:2002px} is not at work. However, contrary to the case studied by EHM, our solution is time dependent. It is natural to ask whether the process of quantum censorship operates only if we impose that the  CFT be static.   More explicitly, one might draw a parallel with the different ways in which a 3+1 dimensional Schwarzschild black hole receives quantum corrections: depending on the choice of boundary conditions, such quantum corrections can be either regular at the horizon and at infinity (in the Unruh state), regular at infinity and time-independent (in the Boulware state) or time-independent and regular at the horizon (in the Hartle-Hawking state). It is tempting to see the dressed conical singularity of EHM as the effect of the backreaction of the CFT in a Boulware-like state (time independent, regular at infinity, singular at the center), whereas the solution discussed in section 3 should be associated to a CFT in a Unruh-like state (regular everywhere but time dependent). Again, it would be interesting to check this behavior on the CFT side of the duality.

Our solutions represent a generalization of the results of EHM, as they depend on one more parameter. For the solutions discussed in section 4, this extra parameter is associated to the acceleration of the brane black hole(s). In the case of the solution of section 3, the extra parameter gives the typical length scale over which the lump of radiation evolves. In general we see that for a given bulk metric we can find a variety of brane induced metrics. We expect such a variety to be present also in the (definitely more interesting and complicated) case of a 3-brane embedded in $(4+1)$-dimensional bulk.

\vspace{1cm}

{\large {\bf \noindent Acknowledgements}}

\smallskip

We thank Roberto Emparan for pointing out an incorrect statement in the first version of this paper and David Kastor for useful discussions. This work has been supported in part by the U.S. National Science Foundation under the grant PHY-0555304.

%
%
\appendix
%
%

\section{No other solutions}

In this appendix we show that the brane embeddings described in the previous sections 3 and 4 are the only possible ones.

Our starting point is the AdS C-metric
\begin{eqnarray}
\nonumber
ds^2&=&\frac{1}{A^2(x-y)^2}\left[-H(y)\,dt^2+\frac{dy^2}{H(y)}+\frac{dx^2}{G(x)}+G(x)\,d\phi^2 \right]\\
H(y)&=&\lambda-k\,y^2+2\,G_4\,mA\,y^3\,\,\,,\,\,\,\, G(x)=1+k\,x^2-2\,G_4\,mA\,x^3\label{AdS c metric}
\end{eqnarray}
with $\lambda >-1$ and $k=-1\,,0\,,+1$. \smallskip

In the following we will be interested in the general embedding of a brane in the above spacetime. We take our brane to be described by the surface $x=\xi(t,y,\phi)$. The unit normal vector is given by
\begin{equation}\label{unit normal for x brane}
n^{\mu}=\frac{A(x-y)}{D_n}\left(\xi_{,t}/H(y)\,,-\xi_{,y} H(y)\,,G(x)\,,-\xi_{,\phi}/G(x) \right)\,,
\end{equation}
where $D_n=\sqrt{-\xi_{,t}^2/H(y)+\xi_{,y}^{2}H(y)+G(x)+\xi_{,\phi}^{2}/G(x)}$.

One can also construct a set of linearly independent vectors tangent to the surface
\begin{eqnarray}
\nonumber
W_{1}^{\mu}&=&\frac{A(x-y)\left(1\,,0\,,\xi_{,t}\,,0\right)}{\sqrt{H(y)-\xi_{,t}^2/G(x)}}\\
\nonumber
W_{2}^{\mu}&=&\frac{A(x-y)\left(0\,,1\,,\xi_{,y}\,,0 \right)}{\sqrt{1/H(y)+\xi_{,y}^{2}/G(x)}}\\
W_{3}^{\mu}&=&\frac{A(x-y)\left(0\,,0\,,\xi_{,\phi}\,,1\right)}{\sqrt{G(x)+\xi_{,\phi}^2/G(x)}}\,.\label{independent frame}
\end{eqnarray} 
The non zero components of the induced metric $h_{ab}$ on the brane are given by $-h_{11}=h_{22}=h_{33}=1$ and
\begin{eqnarray}
\nonumber
h_{12}&=&\frac{\xi_{,t}\xi_{,y}}{G(x)\,\sqrt{H(y)-\xi_{,t}^2/G(x)}\sqrt{1/H(y)+\xi_{,y}^{2}/G(x)}}\\
\nonumber
h_{13}&=&\frac{\xi_{,t}\xi_{,\phi}}{G(x)\,\sqrt{H(y)-\xi_{,t}^2/G(x)}\sqrt{G(x)+\xi_{,\phi}^2/G(x)}}\\
h_{23}&=&\frac{\xi_{,y}\xi_{,\phi}}{G(x)\,\sqrt{1/H(y)+\xi_{,y}^{2}/G(x)}\sqrt{G(x)+\xi_{,\phi}^2/G(x)}}\,.\label{induced metric}
\end{eqnarray}
By direct calculations one can show that the non zero components of the extrinsic curvature $K_{ab}=h^{c}_{(a}h^{d}_{b)}\nabla_{c}n_{d}$ are given by
\begin{eqnarray}
\nonumber
K_{11}&=&A\frac{2G^2H-2\left(G+H\xi_{,y}\right)\xi_{,t}^2 +\left(G'\xi_{,t}^2-2G\xi_{,tt}+GHH'\xi_{,y} \right)(\xi-y)}{2\,D_n(GH-\xi_{,t}^2)}\\
\nonumber
K_{22}&=&-A\frac{2G^2+2H(G+G\xi_{,y}+H\xi_{,y}^2)\xi_{,y}+\left(GH'\xi_{,y}-G'H\xi_{,y}^2+2GH\xi_{,yy} \right)(\xi-y)}{2\,D_n(G+H\xi_{,y}^2)}\\
\nonumber
K_{33}&=&-A\frac{2G^2(G+H\xi_{,y})+2(G+H\xi_{,y})\xi_{,\phi}^2+\left(-G^2G'-3G'\xi_{,\phi}^2+2G\xi_{,\phi\phi}\right)(\xi-y)}{2\,D_n(G^2+\xi_{,\phi}^2)}\\
\nonumber
K_{12}&=&A\frac{-2(1+H\xi_{,y})H\xi_{,t}\xi_{,y}+\left(GH'\xi_{,t}+HG'\xi_{,t}\xi_{,y}-2H\xi_{,ty} \right)(\xi-y)}{2\,GH\,D_n\sqrt{H-\xi_{,t}^2/G}\sqrt{1/H+\xi_{,y}^2/G}}\\
\nonumber
K_{13}&=&-A\frac{(G+H\xi_{,y})\xi_{,t}\xi_{,\phi}+\left(-G'\xi_{,t}\xi_{,\phi}+\xi_{,t\phi}\right)(\xi-y)}{G\,D_n\sqrt{H-\xi_{,t}^2/G}\sqrt{G+\xi_{,\phi}^2/G}}\\
K_{23}&=&-A\frac{(G+H\xi_{,y})\xi_{,t}\xi_{,\phi}+\left(-G'\xi_{,y}\xi_{,\phi}+G\xi_{,y\phi}\right)(\xi-y)}{G\,D_n\sqrt{1/H+\xi_{,y}^2/G}\sqrt{G+\xi_{,\phi}^2/G}}\,,\label{extrinsic curvature}
\end{eqnarray}
where we denote $H'=dH(y)/dy$ and $G'=dG(x)/dx$.

The Isreal junction conditions read
\begin{equation}\label{Isreal JC}
\Delta K_{ab}=-8\,\pi\,G_{4}\left[S_{ab}-\frac{1}{2}S\,h_{ab} \right]\,
\end{equation}
where $\Delta K_{ab}=K_{ab}^{+}-K_{ab}^{-}$ is the jump in the extrinsic curvature, and $S_{ab}$ is the energy momentum tensor localized on the brane. We consider a purely tensional brane, i.e. $S_{ab}=\tau\,h_{ab}$, where $\tau$ is the brane tension. In the following we impose the $Z_2$ symmetry across the brane, and we define the dimensionless parameter $\alpha=2\pi\,G_4\,\tau/A=(1+\delta)/\ell_4\,A$, where $\delta=0$ corresponds to the case of a critical brane.

The junction conditions for our brane read $\Delta K_{ab}=4\pi\, G_4\,\tau\,h_{ab}$, and imply that the ratio $\Delta K_{ab}/h_{ab}$ is a constant. Hence, we can use the conditions $K_{33}h_{23}=K_{23}h_{33}$, $K_{22}h_{12}=K_{12}h_{22}$, $K_{23}h_{12}=K_{12}h_{23}$ and $K_{13}h_{23}=K_{23}h_{13}$, that yield respectively
\begin{eqnarray}
\nonumber
&&\left(\frac{G+\xi_{,\phi}^2/G}{\xi_{,y}^2}\right)_{,\phi}=0\,,\\
\nonumber
&&\left(\frac{GH+H^2\,\xi_{,y}^2}{\xi_{,t}^2}\right)_{,y}=0\,,\\
\nonumber
&&\left(\frac{H\,\xi_{,\phi}^2}{G\,\xi_{,t}^2} \right)_{,y}=0\,,\\
&&\left(\frac{\xi_{,y}}{\xi_{,t}}\right)_{,\phi}=0\,,\label{the four main equations}
\end{eqnarray}
and using the last equation above we can write the first equation in (\ref{the four main equations}) as
\begin{equation}
\label{the 33 k23 equation}
\left(\frac{G+\xi_{,\phi}^2/G}{\xi_{,t}^2}\right)_{,\phi}=0\,.
\end{equation}
We readily integrate this set of equations to obtain
\begin{eqnarray}
\nonumber
&&\frac{G+\xi_{,\phi}^2/G}{\xi_{,t}^2}=F_1(t,y)\,,\\
\nonumber
&&\frac{GH+H^2\,\xi_{,y}^2}{\xi_{,t}^2}=F_2(t,\phi)\,,\\
\nonumber
&&\frac{H\,\xi_{,\phi}^2}{G\,\xi_{,t}^2} =F_3(t,\phi)\,,\\
&&\frac{\xi_{,y}}{\xi_{,t}}=F_4(t,y)\,,\label{integrating the main 4 equations}
\end{eqnarray}
where $F_1$, $F_2$, $F_3$ and $F_4$ are arbitrary functions.
We use the second and the fourth equations in (\ref{integrating the main 4 equations}) to solve for $\xi_{,t}$ and $\xi_{,y}$
to obtain 
\begin{eqnarray}
\nonumber
\xi_{,t}&=&\frac{\sqrt{G(\xi)H(y)}}{\sqrt{F_2(t,\phi)-H^2(y)\,F_{4}^2(t,y)}}\,,\\
\xi_{,y}&=&\frac{F_4(t,y)\sqrt{G(\xi)H(y)}}{\sqrt{F_2(t,\phi)-H^2(y)\,F_{4}^2(t,y)}}\,.\label{the first two eliminations}
\end{eqnarray}
We also can use the first and the third equations in (\ref{integrating the main 4 equations}) to solve for $\xi_{,t}$ and $\xi_{,\phi}$
\begin{eqnarray}
\nonumber
\xi_{,\phi}&=&\frac{G(\xi)\sqrt{F_3(t,y)}}{\sqrt{F_1(t,y)H(y)-F_3(t,\phi)}}\,,\\
\xi_{,t}&=&\frac{\sqrt{G(\xi)H(y)}}{\sqrt{F_1(t,y)H(y)-F_3(t,\phi)}}\,.\label{the second two eliminations}
\end{eqnarray}
Comparing $\xi_{,t}$ in (\ref{the first two eliminations}) and (\ref{the second two eliminations}) we obtain the consistency condition 
\begin{equation}
H^2(y)F_4^2(t,y)+F_1(t,y)H(y)=F_2(t,\phi)+F_3(t,\phi)=E(t)\,
\end{equation}
where $E(t)$ is an arbitrary function of time only. Hence, we eliminate $F_1$ and $F_3$ from the above equations to get
\begin{eqnarray}
\nonumber
\xi_{,t}&=&\frac{\sqrt{G(\xi)H(y)}}{\sqrt{F_2(t,\phi)-H^2(y)\,F_{4}^2(t,y)}}\,,\\
\nonumber
\xi_{,y}&=&\frac{F_4(t,y)\sqrt{G(\xi)H(y)}}{\sqrt{F_2(t,\phi)-H^2(y)\,F_{4}^2(t,y)}}\,,\\
\xi_{,\phi}&=&\frac{G(\xi)\sqrt{E(t)-F_2(t,\phi)}}{\sqrt{F_2(t,\phi)-H^2(y)\,F_{4}^2(t,y)}}\,.\label{the last three main equations}
\end{eqnarray}
Now, we use the equation $K_{33}\,h_{12}=K_{12}\,h_{33}$, which reads,
\begin{equation}
\xi_{,\phi}\left(1+\frac{\xi_{,\phi}^2}{G^2}\right)\left(\frac{H}{\xi_{,t}^2}\right)_{,y}+\xi_{,y}\left(\frac{H}{\xi_{,t}^2}\right)\left(1+\frac{\xi_{,\phi}^2}{G^2}\right)_{,\phi}=0\,,
\end{equation}
along with eq.~(\ref{the 33 k23 equation}) to obtain
\begin{equation}
G\xi_{,\phi}\left(\frac{H}{\xi_{,t}^2}\right)_{,y}=-H\xi_{,y}\left(\frac{G}{\xi_{,t}^2}\right)_{,\phi}\,.\label{simplified k33 k23 equation}
\end{equation}
Substituting $\xi_{,t}$, $\xi_{,\phi}$ and $\xi_{,y}$ from eq. (\ref{the last three main equations}) into eq. (\ref{simplified k33 k23 equation}) we find
\begin{equation}
G^{\prime}H^2F_{4}^3\sqrt{\frac{HG\left(E-F_2\right)}{F_2-H^2F_4^2}}-G\sqrt{E-F_2}\left(H^2F_{4}^2\right)_{,y}=F_4\sqrt{HG}F_{2,\phi}\,.\label{the first of the last two equations}
\end{equation}
In addition using the integrability condition $\xi_{,t\phi}=\xi_{,\phi t}$ we get
\begin{equation}
F_4\sqrt{HG}F_{2,\phi}+F_4\,G^{\prime}\sqrt{HG(E-F_2)(F_2-H^2F_4^2)}=-G\sqrt{E-F_2}\left(H^2F_4^2\right)_{,y}\,.\label{the second of the last two equations}
\end{equation}
By comparing eqs.~(\ref{the first of the last two equations}) and~(\ref{the second of the last two equations}) we finally obtain
\begin{equation}
F_2(t,\phi)\,F_{4}(t,y)\,\sqrt{E(t)-F_2(t,\phi)}=0\,.
\end{equation}
This equation has three possible solutions: $F_2=0$,  $F_2\left(t,\,\phi\right)=E\left(t\right)$ and $F_4=0$. Let us examine them.

Using eq.~(\ref{integrating the main 4 equations}), the condition $F_2=0$ gives $\xi_{,y}^2=-G(\xi)/H(y)$ which forces one of the tangential coordinates on the brane to be light-like, i.e. $W_{2\,\mu}W_2^{\mu}=0$. This situation is not interesting for us and excludes the possibility $F_2=0$.

The second possibility $\sqrt{E(t)-F_2(t,\phi)}=0$, i.e. $F_3=0$,  gives $\xi=\xi(t,y)$. In the following we show that a solution of the form $\xi=\xi(t,y)$ is also forbidden by the junction conditions. We start by using the equations $K_{11}h_{33}=K_{33}h_{11}$, $K_{22}h_{33}=K_{33}h_{22}$  and $K_{12}h_{33}=K_{33}h_{12}$ from which we obtain
\begin{eqnarray}
\nonumber
\xi_{,tt}+H^2(y)\xi_{,yy}&=&0\,,\\
H^{\prime}\xi_{,t}&=&2H\xi_{,yt}\,.\label{the special case yt}
\end{eqnarray} 
The second equation above can be integrated to yield
\begin{equation}
\xi(t,y)=\sqrt{H(y)}\gamma(t)+C(y)\,,
\end{equation}
where $\gamma$ and $C$ are arbitrary functions. Substituting this result into the first equation of (\ref{the special case yt}) we obtain
\begin{equation}\label{nonce}
\frac{d^2\gamma(t)}{dt^2}+\left(HH^{\prime\prime}/2-H^{\prime\,2}/4\right)\gamma(t)+H^{3/2}(y)C^{\prime\prime}(y)=0\,.
\end{equation}
Using $H\left(y\right)=\lambda-ky^2+2\,G_4mA\,y^3$ in the above equation, it is straightforward to see that~(\ref{nonce}) does not have a solution for $m\ne0$, so that  also $F_3=0$ is excluded.

Finally, the condition $F_4$=0 gives $\xi_{,y}=0$ which implies that the possible solution could only be of the form $\xi=\xi(t,\phi)$. However the $K_{12}$ component gives the constraint $\xi_{,t}=0$. Therefore, we are left with $\xi=\xi(\phi)$ as the only possible solution. 

The proof is not complete yet, since we did not cover the case in which the embedding does not depend on the coordinate $x$ (i.e. the case where the brane embedding is described by $y=\psi\left(t,\,\phi\right)$). This case is however easily covered as we observe that the AdS C-metric is invariant under the transformation $t\leftrightarrow i\phi$, $x\leftrightarrow y$, $G\leftrightarrow H$. By using this duality we immediately  see that, if $x=\xi\left(\phi\right)$ solves our system (the solution is discussed in section 4), then also $y=\psi\left(t\right)$ will give a possible embedding (discussed in section 3).

\end{document}